\documentclass[journal=jacsat,manuscript=article]{achemso}
\usepackage[version=3]{mhchem}
\usepackage{xcolor}

\usepackage{titlesec}
\usepackage[toc,page]{appendix}
\usepackage{array}
\usepackage{float}

\newcommand{\bra}[1]{\left\langle #1\right|}
\newcommand{\ket}[1]{\left| #1\right\rangle}

\newcommand{\ev}[1]{\left\langle #1 \right\rangle}

\newcommand{\abs}[1]{|#1|}

\author{Bilal Khalid}
\affiliation[Purdue University]{Department of Physics and Astronomy, Purdue University, West Lafayette, IN-47907, USA}
\author{Shree Hari Sureshbabu}
\affiliation[Purdue University]{Elmore Family School of Electrical and Computer Engineering, Purdue University, West Lafayette, IN-47907, USA}
\author{Arnab Banerjee}
\affiliation[Purdue University]{Department of Physics and Astronomy, Purdue University, West Lafayette, IN-47907, USA}
\author{Sabre Kais}
\email{kais@purdue.edu}
\affiliation[Purdue University]{Department of Chemistry, Department of Physics and Astronomy and Purdue Quantum Science and Engineering Institute, Purdue University, West Lafayette, IN-47907, USA}

\title{Finite-Size Scaling on a Digital Quantum Simulator using Quantum Restricted Boltzmann Machine}

\begin{document}

\begin{abstract}
  The critical point and the critical exponents for a phase transition can be determined using the Finite-Size Scaling (FSS) analysis. This method assumes that the phase transition occurs only in the infinite size limit. However, there has been a lot of interest recently in quantum phase transitions occuring in finite size systems such as a single two-level system interacting with a single bosonic mode e.g. in the Quantum Rabi Model (QRM). Since, these phase transitions occur at a finite system size, the traditional FSS method is rendered inapplicable for these cases. For cases like this, we propose an alternative FSS method in which the truncation of the system is done in the Hilbert space instead of the physical space. This approach has previously been used to calculate the critical parameters for stability and symmetry breaking of electronic structure configurations of  atomic and molecular systems. We calculate the critical point for the quantum phase transition of the QRM using this approach. We also provide a protocol to implement this method on a digital quantum simulator using the Quantum Restricted Boltzmann Machine algorithm. Our work opens up a new direction in the study of quantum phase transitions on quantum devices.
\end{abstract}

\section{Introduction}


A phase transition occurs whenever the thermodynamic functions of a system become non-analytic e.g. as a liquid changes into a gas, the density of the system changes discontinuously. If the phase transition occurs at a finite temperature $T \neq 0$, the transition is called a classical phase transition (CPT) as it is dominated by thermal fluctuations. On the other hand, if the transition occurs by tuning some parameter in the system's Hamiltonian as $T \to 0$, it is called a quantum phase transition (QPT) since it is dominated by quantum fluctuations. A CPT appears only when the system is infinite i.e. $N \to \infty$\cite{goldenfeld}. On the other hand, a QPT doesn't necessarily require $N \to \infty$. Recently there has been a lot of interest in QPTs occurring in finite size light-matter interaction systems\cite{qrm,jc,qrm_anisotropic,qrm_open,rabi_stark,qrm_experiment}.

It has been shown that a QPT occurs in the Quantum Rabi Model (QRM) which describes the interaction of a two-level system with one bosonic field mode\cite{qrm} (see Eq.~\eqref{H_Rabi} for the Hamiltonian). Namely, when the energy separation of the two levels in the system $\Omega$ becomes infinitely large compared to the frequency of the bosonic mode $\omega_0$, the ground state of the Hamiltonian undergoes a phase transition from a normal phase to a superradiant phase as the light-matter coupling exceeds the critical value. Moreover, the ground state of the Jaynes-Cummings model (JCM) which can be obtained from the QRM by performing the rotating-wave approximation has also been shown to exhibit the normal-superradiant phase transition\cite{jc}. Later on, a more general anisotropic QRM in which the rotating and counter-rotating terms can have different coupling strengths was also considered\cite{qrm_anisotropic}. The QRM and JCM are limiting cases of this model. It was shown that the ground state for this more general case also undergoes the normal-superradiant phase transition. The phase transition in QRM has also been demonstrated experimentally using a $^{171}$Yb$^+$ ion in a Paul trap\cite{qrm_experiment}. This experimental demonstration of a phase transition in a single two-level system has incited a lot of interest since this opens up an avenue for studying critical phenomena in controlled, small quantum systems.

In CPTs and some QPTs (which require $N \to \infty$), a finite-size scaling (FSS) analysis can be done to extract the critical point and the critical exponents of the transition\cite{goldenfeld,fss_cpt}. While this procedure is inapplicable to the QPTs discussed above since these phase transitions occur at a finite system size, the phase transitions in these paradigmatic light-matter interaction models occur only in the limit $\Omega/\omega_0 \to \infty$ and FSS analysis can be done in $\Omega/\omega_0$\cite{qrm,jc,qrm_anisotropic} instead. In this paper, however, we propose a different approach to study such phase transitions. We apply the FSS in Hilbert space method\cite{fss1997,fss_yukawa,fss1998,fss_qm,fss2000a,fss2000b} to the QPT in Quantum Rabi Model. In this approach, the truncation of the system is done not in the physical space but in the Hilbert space. The set of basis states spanning the infinite dimensional Hilbert space is truncated to a finite set and the scaling ansatz is employed in terms of the size of this set. This approach has previosuly been developed and applied to a single particle in Yukawa potential\cite{fss_yukawa,fss_qm} and the problem of finding electronic structure critical parameters for atomic and molecular systems\cite{fss1997,fss1998,fss2000a,fss2000b,fss_review}.

In recent years, digital and analog quantum simulators have emerged as a promising platform for the simulation of quantum phenomena. Quantum simulators have already been used to study phase transitions using the method of partition function zeros\cite{pf_zeros} and the Kibble-Zurek mechanism\cite{kz_lukin,moore_pt}. In this paper, we present a protocol to implement the finite-size scaling method on a digital quantum simulator. We use the Quantum Restricted Boltzmann Machine (QRBM) algorithm to find the critical point of the Quantum Rabi model.

This paper is organized as follows. In Sec.~\ref{theory}, we explain the theory of Quantum Rabi Model, Finite-Size Scaling and the Quantum Restricted Boltzmann Machine. In Sec.~\ref{results}, we present our results obtained using the exact diagonalization method and QRBM. Finally in Sec.~\ref{discussion}, we discuss our results and future prospects of studying quantum phase transitions on quantum devices. 

\section{Theory}
\label{theory}

\subsection{Quantum Rabi Model}

The QRM describes a two-level system interacting with a bosonic field mode. The Hamiltonian is\cite{qrm},
\begin{equation}
    H_{Rabi} = \frac{\Omega}{2} \sigma_z + \omega_0 a^\dagger a - \lambda \sigma_x (a + a^\dagger)
\label{H_Rabi}
\end{equation}
where we've chosen $\hbar = 1$. $\Omega$ is the energy separation between the two levels in the system, $\omega_0$ is the frequency of the bosonic mode and $\lambda$ is the system-environment coupling strength. The parity operator $\Pi = e^{i \pi (a^\dagger a + \ket{\uparrow}\bra{\uparrow})}$ commutes with $H_{Rabi}$. So, $H_{Rabi}$ has a $Z_2$ symmetry.

This model has a critical point at $g=2\lambda/\sqrt{\omega_0 \Omega}=g_c=1$ in the limit $\Omega/\omega_0 \to \infty$\cite{qrm}. For $g<1$, the system is in the \textit{normal phase} and the ground state is $\ket{\phi^0_{np} (g)} = \mathcal{S}[r_{np}(g)] \ket{0} \ket{\downarrow}$ where $\mathcal{S}[x] = \exp [\frac{x}{2} (a^{\dagger 2}-a^2)]$ and $r_{np}(g) = -\frac{1}{4} \ln (1 - g^2)$. The rescaled ground state energy and photon number are $e_G(g)=\frac{\omega_0}{\Omega} \ev{H_{Rabi}}=-\omega_0/2$ and $n_G(g)=\frac{\omega_0}{\Omega} \ev{a^\dagger a}=0$ respectively. For $g>1$, the system is in a \textit{superradiant phase} and the ground state is two-fold degenerate, $\ket{\phi^0_{sp} (g)} = \mathcal{D}[\pm \alpha_g] \mathcal{S}[r_{sp}(g)] \ket{0} \ket{\downarrow^{\pm}}$ here $r_{sp}(g) = -\frac{1}{4} \ln (1 - g^{-4})$ and $\mathcal{D}[\alpha]=\exp [\alpha (a^\dagger -a)]$. $\ket{\downarrow^{\pm}}$ is the negative eigenvalue eigenstate of $\frac{1}{2 g^2} \sigma_z \pm \frac{2 \lambda \alpha_g}{g^2 \Omega} \sigma_x$ where $\alpha_g = \sqrt{\frac{\Omega}{4 g^2 \omega_0} (g^4-1)}$. The rescaled ground state energy and photon number are $e_G(g)=\frac{\omega_0}{\Omega} \ev{H_{Rabi}}=-\omega_0 (g^2+g^{-2})/4$ and $n_G(g)=\frac{\omega_0}{\Omega} \ev{a^\dagger a}=(g^2-g^{-2})/4$ respectively.

\begin{figure}[h]
\centering
\includegraphics[width=1.04\linewidth]{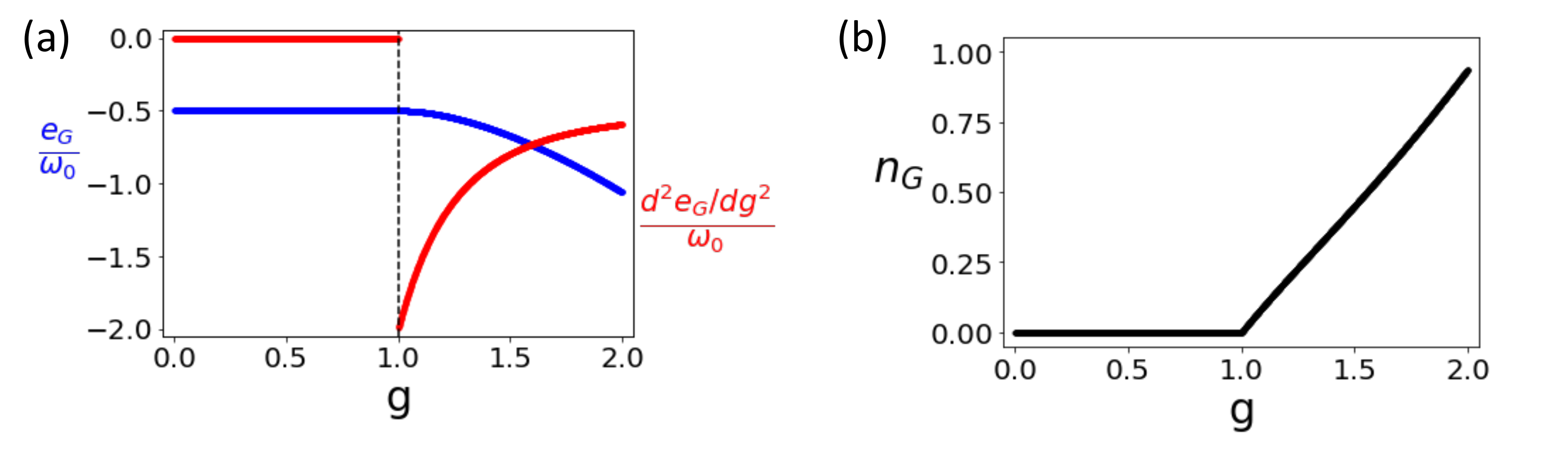}
\caption{\textbf{Phase Transition in Quantum Rabi Model.} (a) The rescaled ground state energy $e_G/\omega_0 = \ev{H_{Rabi}}/\Omega$ and $(d^2 e_G/d g^2)/\omega_0$ as functions of $g$. The discontinuity in $(d^2 e_G/d g^2)/\omega_0$ at $g=g_c=1$ indicates a countinuous phase transition. (b) The order parameter $n_G=\frac{\omega_0}{\Omega} \ev{a^\dagger a}$ as a function of $g$. $n_G$ becomes non-zero when the $Z_2$ symmetry is spontaneously broken at $g>g_c=1$.}
\label{qrm}
\end{figure}

As shown in Fig.~\ref{qrm}(a) and (b), $d^2 e_G/dg^2$ is discontinuous at $g=g_c=1$, indicating a continuous phase transition and $n_G=\frac{\omega_0}{\Omega} \ev{a^\dagger a}$ is an order parameter for this phase transition. In the \textit{normal phase}, $n_G$ is zero whereas in the \textit{superradiant phase}, $Z_2$ symmetry is spontaneously broken and $n_G$ becomes non-zero.

We can also write effective low-energy Hamiltonians in both the \textit{normal} and the \textit{superradiant phases}. For $g<1$, $H_{Rabi}$ can be reduced to the following effective Hamiltonian\cite{qrm},
\begin{equation}
    H_{np} = \omega_0 a^\dagger a - \frac{\omega_0 g^2}{4} (a + a^\dagger)^2 - \frac{\Omega}{2}.
\label{H_np}
\end{equation}
The system's degrees of freedom have been removed by projecting to $\ket{\downarrow}\bra{\downarrow}$, since this is a low energy description. Similarly, for $g>1$ the effective Hamiltonian can be written as\cite{qrm},
\begin{equation}
    H_{sp} = \omega_0 a^\dagger a - \frac{\omega_0}{4 g^4} (a + a^\dagger)^2 - \frac{\Omega}{2} (g^2 + g^{-2}),
\label{H_sp}
\end{equation}
where this time around the Hamiltonian has been projected along $\ket{\downarrow^{\pm}}\bra{\downarrow^{\pm}}$. In Sec.~\ref{results}, we'll use $H_{np}$ and $H_{sp}$ to find the critical point of the model.

\subsection{Finite-Size Scaling}

The FSS method is widely used to determine the critical points and the critical exponents in phase transitions\cite{goldenfeld}. To demonstrate the method, consider that we have an infinite $2d$ system that undergoes a classical phase transition at a critical temperature $T=T_c$\cite{fss_cpt}. Suppose $Q$ is a quantity that becomes singular at $T=T_c$ with some power law behavior
\begin{equation}
    Q_\infty (T) \sim \abs{T-T_c}^{- \omega}.
\label{Q}
\end{equation}
We can also think of this system as an infinite collection of infinite stripes, where the stripes are infinitely extended along one direction and stacked along the perpendicular direction. Now suppose there are only an $N$ number of stripes. If $N$ is finite, $Q$ should be regular at $T=T_c$ since finite systems cannot have non-analyticities at $T \neq 0$. The singularity at $T=T_c$ should appear only when $N \to \infty$. The finite size scaling hypothesis assumes the existence of a scaling function $F_Q$ such that
\begin{equation}
    Q_N(T) \simeq Q_{\infty} (T) F_Q(N/\xi_{\infty}(T)),
\label{scaling}
\end{equation}
where $Q_N$ is the observable $Q$ for a system with $N$ stripes and $Q_\infty$ corresponds to the system in the thermodynamic limit. $\xi_\infty$ is the correlation length for the infinite system. Eq.~\eqref{scaling} is valid when $N$ is large. The correlation length also diverges as a power law near the critical point,
\begin{equation}
    \xi_\infty (T) \sim \abs{T-T_c}^{-\nu}.
\label{correlation}
\end{equation}
Substituting Eq.~\eqref{Q} and \eqref{correlation} in Eq.~\eqref{scaling},
\begin{equation}
    Q_N(T) \simeq \abs{T-T_c}^{-\omega} F_Q (N \abs{T-T_c}^\nu).
\label{scaling1}
\end{equation}
Since $Q_N(T)$ should be regular at $T=T_c$, the scaling function should cancel the divergence due to $\abs{T-T_c}^{-\omega}$. Therefore, the scaling function should be of the form $F_Q(x) \sim x^{\omega/\nu}$ as $x \to 0$. We should then have,
\begin{equation}
    Q_N(T_c) \sim N^{\omega/\nu}.
\end{equation}
If we define a function $\Delta_Q (T;N,N')$ such that
\begin{equation}
    \Delta_Q (T;N,N') = \frac{\log (Q_N(T)/Q_{N'}(T))}{\log (N/N')},
\label{delta}
\end{equation}
then the value of this function at $T=T_c$, $\Delta_Q (T_c;N,N') \simeq \omega/\nu$ is independent of $N$ and $N'$. Therefore, for three different values $N$, $N'$ and $N''$, the curves $\Delta_Q (T;N,N')$ and $\Delta_Q (T;N',N'')$ will intersect at the critical point $T=T_c$. This is how we can locate the critical point using the finite size scaling hypothesis.

We can also find the critical exponents $\omega$ and $\nu$. Noting from Eq.~\eqref{Q} that
\begin{equation}
    \frac{\partial Q_\infty (T)}{\partial T} \sim \abs{T-T_c}^{-(\omega+1)}.
\end{equation}
Therefore, we should have $\Delta_{\partial Q/\partial T} (T_c;N,N') \simeq (\omega+1)/\nu$. Define a new function $\Gamma_\omega (T;N,N')$ such that
\begin{equation}
    \Gamma_\omega (T;N,N') = \frac{\Delta_Q(T;N,N')}{\Delta_{\partial Q/\partial T}(T;N,N') - \Delta_Q(T;N,N')}.
\label{gamma}
\end{equation}
The value of this function at the critical point $\Gamma_\omega (T_c;N,N') \simeq \omega$ is independent of $N$ and $N'$ and gives us the critical exponent $\omega$. Then $\nu$ can be determined using
\begin{equation}
    \nu \simeq \frac{\omega}{\Delta_Q (T_c;N,N')}.
\end{equation}

As we've already stated in the \textit{Introduction}, this method cannot be used for the kinds of phase transitions we are interested in which occur at a finite system size. However, for such cases we can consider an extension of the approach discussed above\cite{fss1997,fss_yukawa,fss1998,fss_qm,fss2000a,fss2000b,fss_review}. In this extended approach, instead of truncating the system in the physical space, the system is truncated in the Hilbert space\cite{fss_review}. The FSS ansatz looks exactly the same except that $N$ now represents the size of the set of basis states which spans the truncated Hilbert space\cite{fss_review}. Moreover, the temperature $T$ will be replaced by the parameter $g$ which is being tuned across the critical point. This approach has been shown by Kais and co-workers to work in the case of a particle in Yukawa potential\cite{fss_yukawa,fss_qm} and the calculation of electronic structure critical parameters for atomic and molecular systems\cite{fss1997,fss1998,fss2000a,fss2000b,fss_review}.

\subsection{Quantum Restricted Boltzmann Machine}

Solving quantum many-body problem accurately has been a taxing numerical problem since the size of the wavefunction scales exponentially. The idea of taking advantage of the aspects of Machine Learning (ML) related to dimensionality reduction and feature extraction to capture the most relevant information came from the work by Carleo and Troyer \cite{carleo2017solving}, which introduced the idea of representing the many-body wavefunction in terms of an Artificial Neural Network (ANN) to solve for the ground states and time evolution for spin models. A Restricted Boltzmann Machine (RBM) was chosen as the architecture of this ANN. An RBM consists of a visible layer and a hidden layer with each neuron in the visible layer connected to all neurons in the hidden layer but the neurons within a layer are not connected to each other. The quantum state is  $\psi$ expanded in the basis $\ket{x}$:
\begin{equation}
	\ket{\psi} = \sum{\psi(x) \ket{x}}
\end{equation}

The Neural Network Quantum State (NQS) describes the wavefunction $\psi(x)$ to be written as $\psi(x;\theta)$, where $\theta$ represents the parameters of the RBM. $\psi(x; \theta)$ is now written in terms of the probability distribution that is obtained from the RBM as follows:

\begin{equation}
	\psi(x; \theta) \propto \sum_{\{h\}}e^{\frac{1}{2} \sum_i a_i \sigma_i^z + \sum_j b_j h_j + \sum_{ij} w_{ij} \sigma_i^z h_j}
\end{equation}
where, $\sigma_i^z$ is the Pauli z operator at $i^{th}$ site, $\sigma_i^z$ and $h_j$ take values $\{+1, -1\}$, $\theta = \{a_i, b_j, w_{ij}\}$ are the trainable bias and weight parameters of the RBM. Using stochastic optimization, the energy $E(\theta)$ is minimized. 

This work was extended to obtain the ground states of the Bose-Hubbard model \cite{saito2017solving} and for the application of quantum state tomography \cite{torlai2018neural}. 

With the rapid developments in the domains of ML and Quantum Computing (QC), the appetite for integrating ideas in both of these areas has been growing considerably. The last decade has seen a surge in the application of classical ML for quantum matter, wherein these methods have been adopted to benchmark, estimate and study the properties of quantum matter \cite{carrasquilla2017machine, deng2017machine, butler2018machine, carleo2019machine}, with recently showing provable classification efficiency in classifying quantum states of matter \cite{huang2021provably}. The protocols and algorithms related to ML implementable on a quantum system so called Quantum machine Learning \cite{Biamonte_2017} is expected to have the potential of changing the course of fundamental scientific research \cite{sajjan2021quantum} along with industrial pursuit. 

In lieu of today’s Noisy Intermediate Scale Quantum (NISQ) devices, the ideas which utilize both classical and quantum resources, such that the part of the problem which has an exponential scaling is implemented on the quantum platform while the rest are dealt with classically, are being carefully investigated for various applications. Such algorithms are known as classical-quantum hybrid algorithms. In the work by Xia and Kais \cite{Xia_2018}, a modified RBM with three layers was introduced, the third layer to account for the sign of the wavefunction, to solve for the ground state energies of molecules. Now, the parametrized wavefunction $\psi(x; \theta)$ is written as a function of $P(x)$ along with a sign function $s(x)$:
\begin{align}
P({\bf x}) &= \frac{\sum_{\{h\}}e^{\sum_{i}a_i\sigma^z_i + \sum_{j}b_j h_j + \sum_{ij}w_{ij}\sigma^z_i h_j}}
{\sum_{\bf x'}\sum_{\{h\}}e^{\sum_{i}a_i\sigma^{z'}_i + \sum_{j}b_j h_j + \sum_{ij}w_{ij}\sigma^{z'}_i h_j}}\\[11pt]
s(\bf x) &= \tanh\left[(c + \sum_{i}d_i\sigma_i) \right]
\end{align}

The wavefunction ansatz in terms of the RBM can be expressed as\cite{Xia_2018}:
\begin{equation}
	\ket{\psi} = \sum_x \sqrt{P(x)} s(x) \ket{x}
\end{equation}

\begin{figure}[H]
    \centering
    \includegraphics[width=0.75\textwidth]{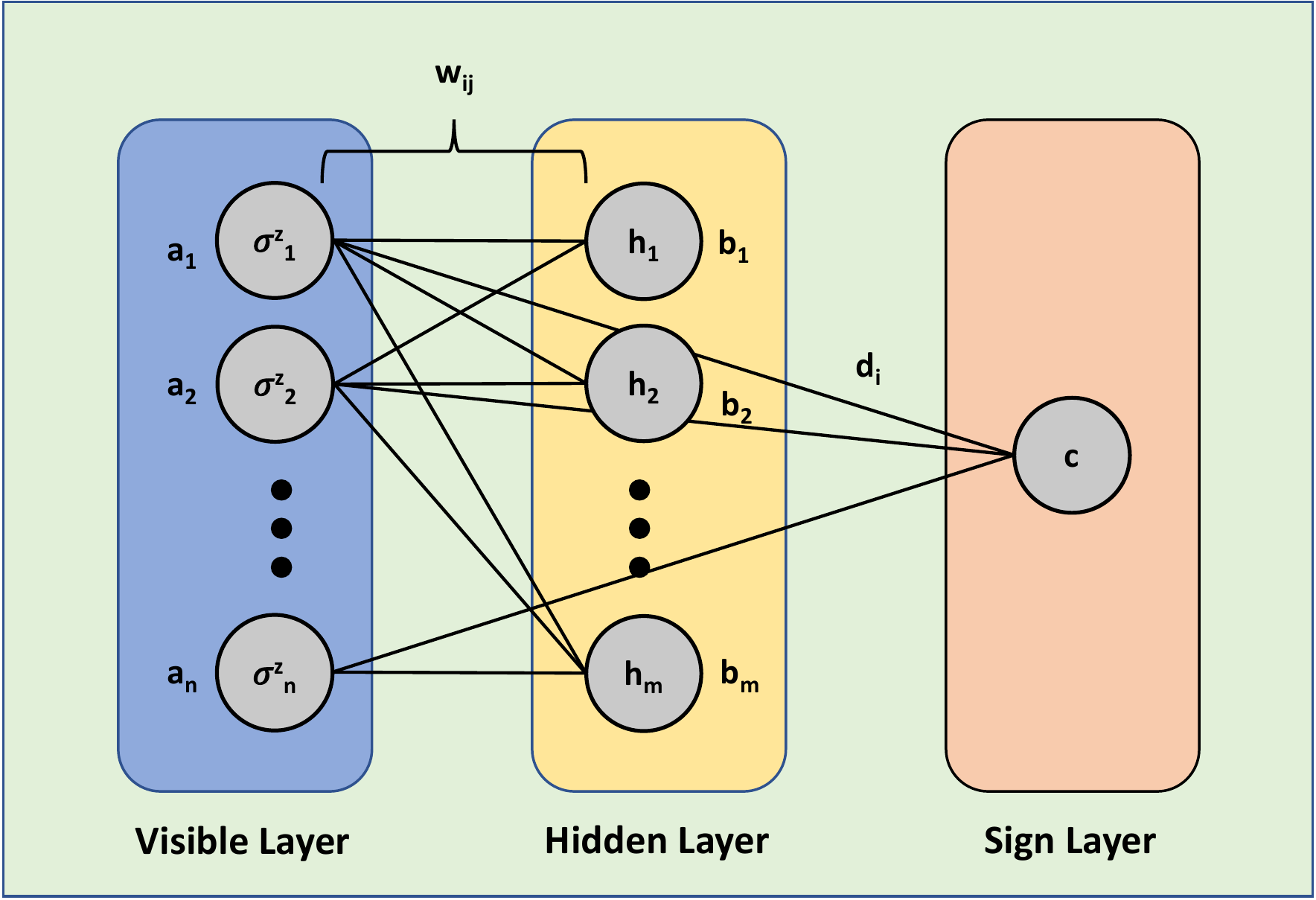}
    \caption{\textbf{Restricted Boltzmann Machine architecture.} The first layer is the visible layer with bias parameters denoted by $a_i$. The second layer is the hidden layer with bias parameters denoted by $b_j$. The third layer is the sign layer with bias parameters denoted by $c$. The weights associated with the connections between the visible neurons and the hidden neurons are designated by $w_{ij}$. The weights associated with the connections between the visible neurons and the neuron of the sign layer are designated by $d_{i}$.}
    \label{QRBM}
\end{figure}

\begin{figure}[H]
    \centering
    \includegraphics[width=0.85\textwidth]{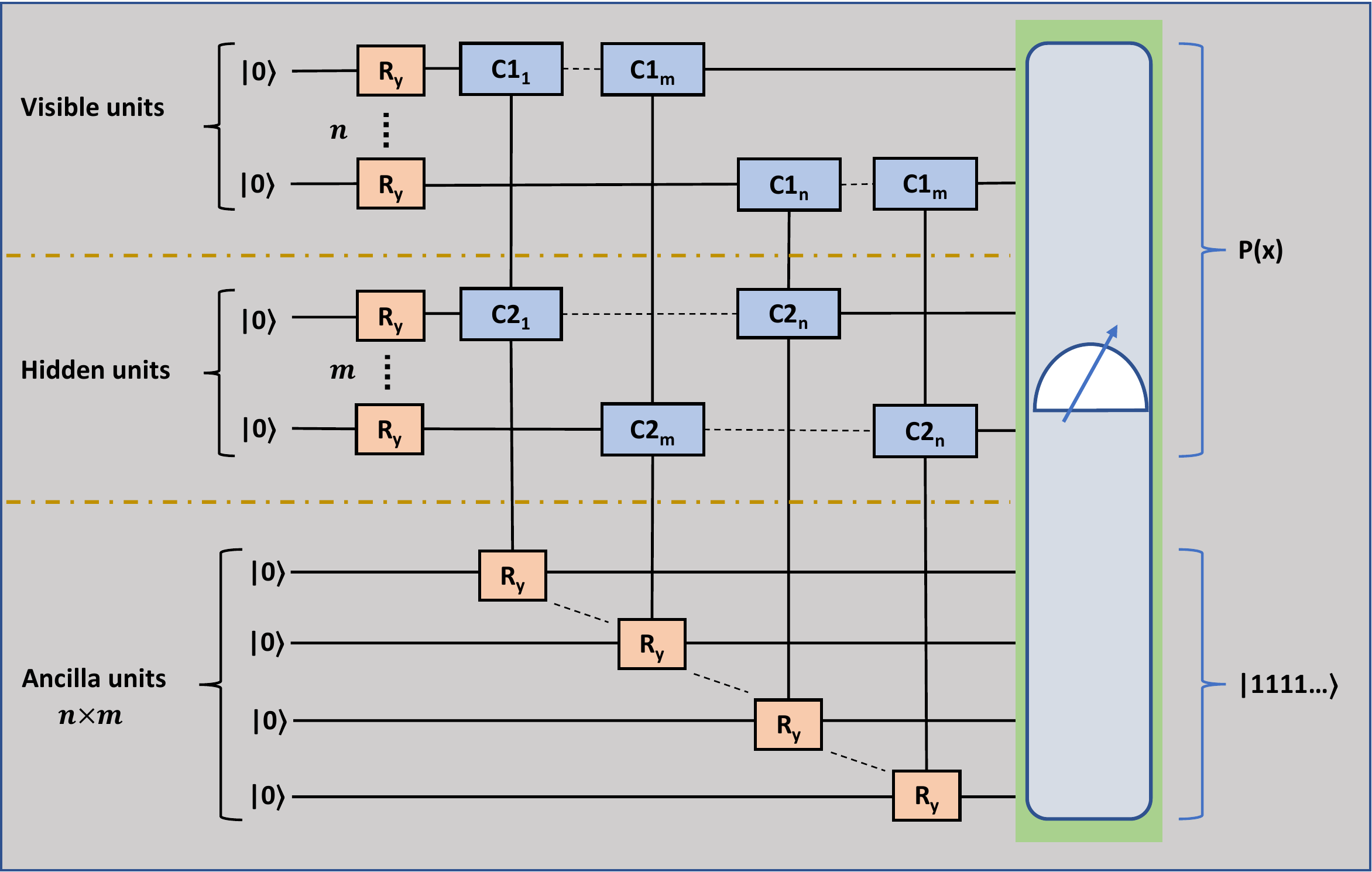}
    \caption{\textbf{The quantum circuit to sample the Gibbs distribution.} $n$ is the number of qubits belonging to the visible layer and $m$ is the number of qubits belonging to the hidden layer. There are $m \times n$ ancillary qubits.}
    \label{gibbs_circuit}
\end{figure}

A quantum circuit comprising of a single-qubit ($R_y$) and multi-qubit y-rotation gates ($C1-C2-R_y$) are employed, to sample the Gibbs distribution. The utilization of $R_y$ gates cater to the bias parameter of visible and hidden layers part of the distribution, while  $C1-C2-R_y$ gates tend to the weights part of the distribution. In the work by Sureshbabu et al. \cite{sureshbabu2021implementation}, the implementation of such a circuit on IBM-Q devices were shown, wherein a new ancillary qubit is introduced to store the value corresponding to every $C1-C2-R_y$ gate (Fig.~\ref{gibbs_circuit}). The term $n$ denotes the number of visible qubits and $m$ denotes the number of hidden units. In this formalism, the number of ancillary qubits required are $n \times m$. Starting all the qubits from a $\ket{0}$, the $R_y$ and $C1-C2-R_y$ rotations are performed, and a measurement is performed on all the qubits. If all the ancillary qubits are in $\ket{1}$, then the sampling is deemed successful and the states corresponding to the first $m + n$ qubits provide the distribution $P(x)$. The joint probability distribution defined over the parameters of the circuit $\theta = \{a, b, w\}$ and a set of $y = \{\sigma^z, h\}$ is given by:
\begin{equation}
	P(y, \theta) = \frac{e^{\sum_{i}a_i\sigma^z_i + \sum_{j}b_j h_j + \sum_{ij}w_{ij}\sigma^z_i h_j}}
{\sum_{\{y\}}e^{\sum_{i}a_i\sigma^{z'}_i + \sum_{j}b_j h_j + \sum_{ij}w_{ij}\sigma^{z'}_i h_j}}
\end{equation}
The probability of successful sampling can be improved by rewriting the distribution $P(y, \theta)$ as $Q(y, \theta)$ and setting $k=max(1, \frac{|w_{ij}|}{2}) $\cite{Xia_2018, doi:10.1021/jacs.1c06246}:
\begin{equation}
	Q(y, \theta) = \frac{e^{\frac{1}{k}(\sum_{i}a_i\sigma^z_i + \sum_{j}b_j h_j + \sum_{ij}w_{ij}\sigma^z_i h_j)}}
{\sum_{\{y\}}e^{\frac{1}{k}(\sum_{i}a_i\sigma^{z'}_i + \sum_{j}b_j h_j + \sum_{ij}w_{ij}\sigma^{z'}_i h_j)}}
\end{equation}

Firstly, the QRBM is implemented classically, i.e, the quantum circuit is simulated on a classical computer. This execution caters to the ideal results that can be obtained through the QRBM algorithm.
Then, the quantum circuit is implemented on the Digital Quantum Simulator, the \textit{qasm} simulation backend. This simulator is part of the high-performance simulators from IBM-Q. The circuit is realized using IBM's Quantum Information Software Toolkit titled Qiskit \cite{aleksandrowicz2019qiskit}. Though no noise model was utilized, as a result of finite sampling, statistical fluctuations in the values of probabilities in observing the circuit in the measurement basis, are present in the obtained results.

Having obtained the distribution $Q(y, \theta)$, the probabilities are raised to the power of $k$, to get $P(y, \theta)$. Following this, the sign function is computed classically, thereby calculating $\ket{\psi}$. Then, the expectation value for the Hamiltonian H [$\bra{\Psi}H\ket{\Psi}$] is computed to get the energy, which is minimized using gradient descent to obtain the ground state eigenenergy of H. 

The resource requirements demanded by this algorithm are quadratic. The number of qubits required are $(m+n)$ to encode the visible and hidden nodes, and $(m \times n)$ to account for the ancillary qubits. Hence, the number of qubits scales as $O(mn)$. The number of $R_y$ gates required are $(m+n)$ and the number of $C1-C2-R_y$ gates required are $(m \times n)$. In addition, each $C1-C2-R_y$ gate requires $6n$ $X$-gates  to account for all the states spanned by the control qubits. Therefore, the number of gates required also scales as $O(mn)$. The time complexity for obtaining the ground states or minimum eigenvalues of a given matrix has a time complexity of $j^3$, with $j$ being the dimension of the column space for the given matrix. 

\section{Results}
\label{results}

\subsection{Exact Diagonalization}

In this section, we demonstrate the calculation of the critical point of the Quantum Rabi model using the Finite-Size Scaling method. As discussed before, the phase transition in QRM occurs only in the limit $\Omega/\omega_0 \to \infty$. This limit is not straightforward to implement in $H_{Rabi}$ given in Eq.~\eqref{H_Rabi}. Instead, we have considered the effective low-energy Hamiltonians $H_{np}$ and $H_{sp}$ given in Eq.~\eqref{H_np} and \eqref{H_sp} respectively. In $H_{np}$ and $H_{sp}$, $\Omega$ is involved only in a constant term which can be removed from the Hamiltonians and the limit $\Omega/\omega_0 \to \infty$ can then be easily imposed.

\begin{figure}[H]
\centering
\includegraphics[width=1.1\linewidth]{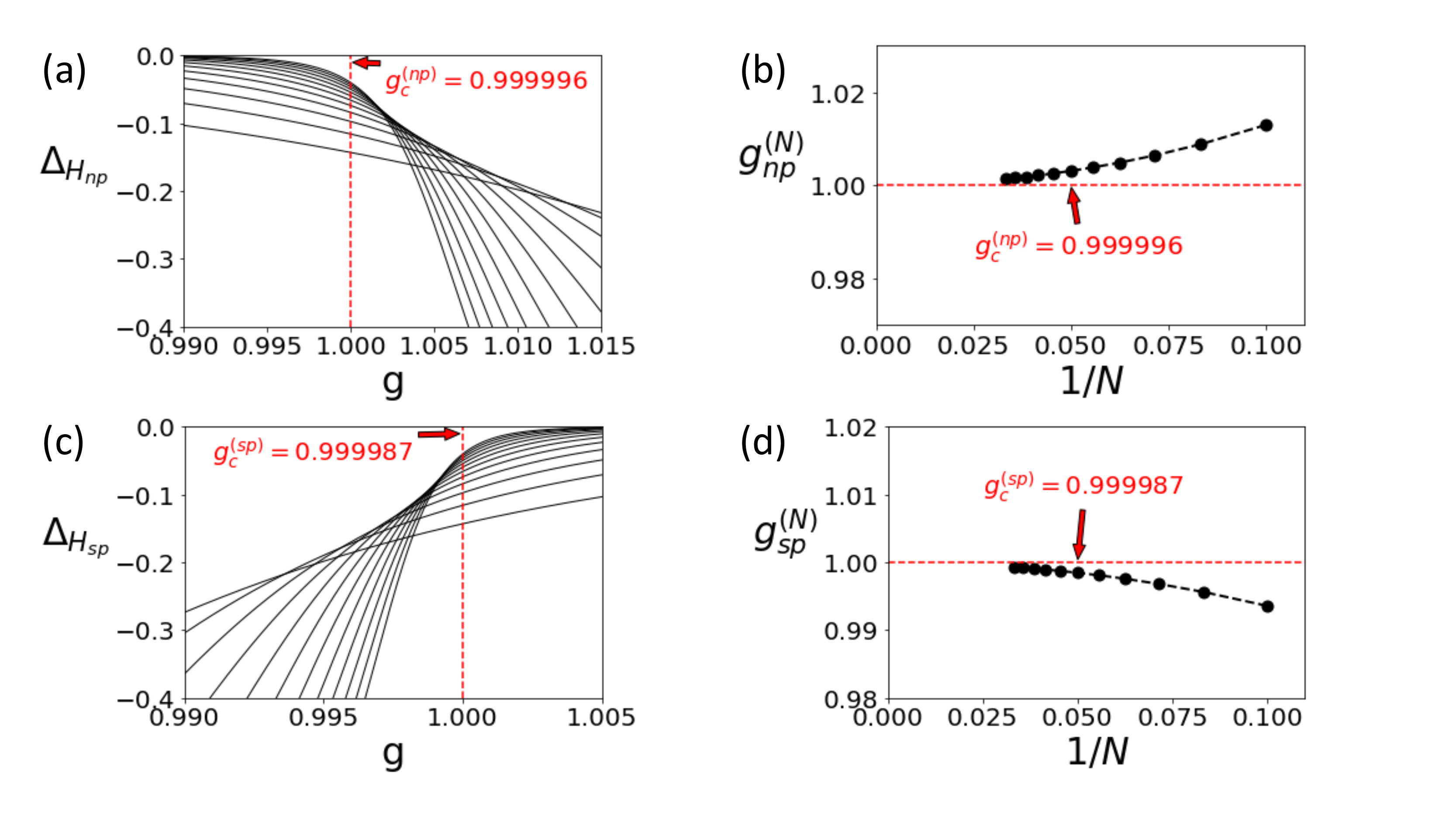}
\caption{\textbf{Finite-Size Scaling for Quantum Rabi model.} We used $N=8,10,\ldots,32$. (a) Graphs of $\Delta_{H_{np}}(g;8,10),\: \Delta_{H_{np}}(g;10,12), \ldots, \: \Delta_{H_{np}}(g;30,32)$ as a function of $g$. (b) Intersection points $g^{(N)}_{np}$ where $\Delta_{H_{np}}(g^{(N)}_{np};N-4,N-2)=\Delta_{H_{np}}(g^{(N)}_{np};N-2,N)$, as a function of $1/N$. As $N \to \infty$, $g^{(N)}_{np} \to 0.999996$. So, $g_c^{(np)} = 0.999996$. (c) Graphs of $\Delta_{H_{sp}}(g;8,10),\: \Delta_{H_{sp}}(g;10,12), \ldots, \: \Delta_{H_{sp}}(g;30,32)$ as a function of $g$. (d) Intersection points $g^{(N)}_{sp}$ where $\Delta_{H_{sp}}(g^{(N)}_{sp};N-4,N-2)=\Delta_{H_{sp}}(g^{(N)}_{sp};N-2,N)$, as a function of $1/N$. As $N \to \infty$, $g^{(N)}_{sp} \to 0.999987$. So, $g_c^{(sp)} = 0.999987$.}
\label{qrm_fss}
\end{figure}

In $H_{np}$ and $H_{sp}$, the degrees of freedom of the two-level system have been traced out and the only degrees of freedom we have are those of the bosonic mode. Let's first consider the \textit{normal phase} Hamiltonian $H_{np}$. The Hilbert space for this Hamiltonian is spanned by the familiar harmonic oscillator number states $\{\ket{0},\ket{1},\ket{2},\ldots\}$. We can truncate the full Hilbert space to an N-dimensional Hilbert space spanned by $\{\ket{0},\ket{1},\ldots,\ket{N-1}\}$ to apply the finite-size scaling analysis. In this restricted Hilbert space, the matrix form of $H_{np}^{(N)}$ can be found by using $a \ket{m} = \sqrt{m} \ket{m-1}$ and $a^\dagger \ket{m} = \sqrt{m+1} \ket{m+1}$. Once we have the matrix form, we can then use the exact diagonalization method to find the ground state of $H_{np}^{(N)}$ with energy $E_{np}^{(N)}$.

Consider the scaling law for the ground state energy in the vicinity of the critical point $g=g_c$,
\begin{equation}
    E(g) \sim \abs{g-g_c}^\alpha.
\label{E}
\end{equation}
Here $E$ is the ground state energy. We slightly modify the formula in Eq.~\eqref{delta} to take into account the difference in the signs of the exponents in Eq.~\eqref{Q} and \eqref{E}. The new formula with $Q=E$ is,
\begin{equation}
    \Delta_{H_{np}} (g;N,N') = \frac{\log (E_{np}^{(N)}(g)/E_{np}^{(N')}(g))}{\log (N'/N)},
\label{delta_hs}
\end{equation}
We plot the curves $\Delta_{H_{np}} (g;N,N+2)$ for $N=8,10,\ldots,30$ in Fig.~\ref{qrm_fss}(a). We then plot the intersection points $g^{(N)}_{np}$ of the curves $\Delta_{H_{np}} (g;N-4,N-2)$ and $\Delta_{H_{np}} (g;N-2,N)$ as a function of $N$ as shown in Fig.~\ref{qrm_fss}(b). To find the limit of $g^{(N)}_{np}$ as $N \to \infty$, we used the Bulirsch-Stoer algorithm (see Appendix~\ref{bsa}). The limit was calculated to be $g^{(N)}_{np} \to 0.999996$. So $g_c^{(np)}=0.999996$.

In a similar way, we then consider $H_{sp}$. The curves $\Delta_{H_{sp}} (g;N,N+2)$ are plotted in Fig.~\ref{qrm_fss}(c) for $N=8,10,\ldots,30$ and the intersection points $g^{(N)}_{sp}$ are plotted in Fig.~\ref{qrm_fss}(d) as a function of $N$. In this case, the extrapolation to $N \to \infty$ gives the critical value $g_c^{(sp)}=0.999987$. Both the calculated values of $g_c^{(np)}$ and $g_c^{(sp)}$ are very close to the exact value $g_c=1$.

\subsection{Quantum Restricted Boltzmann Machine}

\begin{figure}[H]
    \centering
    \includegraphics[width=1\textwidth]{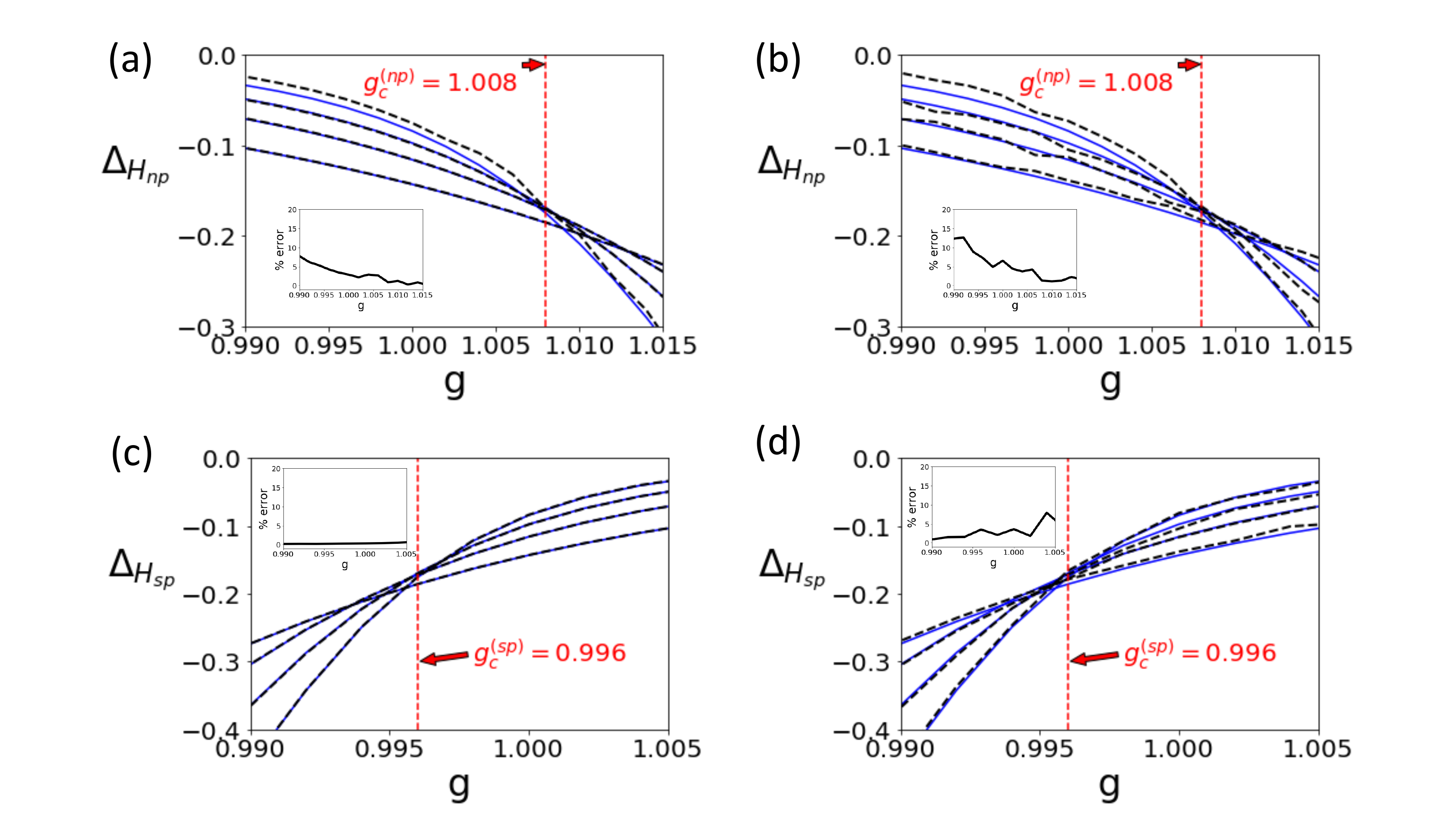}
    \caption{\textbf{QRBM Implementation of FSS for QRM.} The light blue line represents results obtained from exact diagonalization and dashed black line represents QRBM results. (a) Classical implementation of QRBM corresponding to normal phase, graphs of $\Delta_{H_{np}}(g;8,10),\: \Delta_{H_{np}}(g;10,12), \ldots, \: \Delta_{H_{np}}(g;14,16)$ as a function of $g$. (b) QRBM implemented on \textit{qasm} simulator corresponding to normal phase, graphs of $\Delta_{H_{np}}(g;8,10),\: \Delta_{H_{np}}(g;10,12), \ldots, \: \Delta_{H_{np}}(g;14,16)$ as a function of $g$. The $g_c^{(np)}$ in both the cases is calculated to be 1.008. (c) Classical implementation of QRBM corresponding to superradiant phase, graphs of $\Delta_{H_{sp}}(g;8,10),\: \Delta_{H_{sp}}(g;10,12), \ldots, \: \Delta_{H_{sp}}(g;14,16)$ as a function of $g$. (b) QRBM implemented on \textit{qasm} simulator corresponding to superradiant phase, graphs of $\Delta_{H_{sp}}(g;8,10),\: \Delta_{H_{sp}}(g;10,12), \ldots, \: \Delta_{H_{sp}}(g;14,16)$ as a function of $g$. The $g_c^{(sp)}$ in both the cases is calculated to be 0.996. The inset plots display the mean percentage error between the exact diagonalization results and QRBM results.} 
    \label{QRBM_results}
\end{figure}

Now we illustrate the implementation of the FSS method using the QRBM algorithm. The results are shown in Fig.~\ref{QRBM_results}. Fig.~\ref{QRBM_results}(a) and Fig.~\ref{QRBM_results}(c) show the results for $H_{np}$ and $H_{sp}$ using the classical implementation of the algorithm respectively. Whereas, Fig.~\ref{QRBM_results}(b) and Fig.~\ref{QRBM_results}(d) correspond to the results for $H_{np}$ and $H_{sp}$ when the algorithm is implemented using the \textit{qasm} simulator from IBM-Q respectively. The QRBM algorithm is run for $N=8, 10, 12, 14, 16$. 

For the case of N=8, the number of qubits associated with the visible nodes equal 3, the number of qubits associated with the hidden nodes equal 3, and 9 ancillary qubits were used. The quantum circuit consists of 6 $R_y$ gates associated with the bias parameters, 9 $C1-C2-R_y$ gates associated with the weights. Since, each $C1-C2-R_y$ gate requires 6 $X$-gates, a total of 54 $X$-gates were used. 
For the case of N=10,..,16, the number of qubits associated with the visible nodes equal 4, the number of qubits associated with the hidden nodes equal 4, and 16 ancillary qubits were used. 
The quantum circuit consists of 8 $R_y$ gates associated with the bias parameters, 16 $C1-C2-R_y$ gates associated with the weights. Since, each $C1-C2-R_y$ gate requires 6 $X$-gates, a total of 96 $X$-gates were used. 

Starting from random initialization, all parameters are updated via gradient descent. A learning rate of 0.01 was chosen and the algorithm is run for around 30,000 iterations. In order to assist with the convergence to the minimum eigenenergies, warm starting is employed. The method of warm starting is essentially initializing the parameters of the current point with the parameters of a previously converged point of calculation, which helps in avoiding the convergence to a local minima.

The black curves plotted in the insets in Fig.~\ref{QRBM_results} represent the deviation of the QRBM results (black dashed curves) from the exact diagonalization results (blue solid curves). They were calculated using the average of the quantity $\abs{\Delta^{(ED)}(g) - \Delta^{(QRBM)}(g)/\Delta^{(ED)}(g)} \times 100$ over all the four curves. An enlarged version of the error plots can be found in the \textit{Supporting Information} section. For each case the overall error close to $g = 1.000$ is not more than $\sim 5\%$ which implies convergence to the right result. Moreover, for the case of $H_{sp}$, we notice that the error is very small for the classical implementation i.e. $\sim <1\%$ throughout the range of the graph. An astoundingly low error for this particular case shows that the QRBM method is particularly effective in finding the correct ground state for the case of $H_{sp}$. Overall this result also underscores the fact that QRBM can be more effective for certain forms of the Hamiltonian over others, such as in this case it was quite effective for $H_{sp}$ even with a relatively small number of qubits used in the hidden layer.

The critical point using $H_{np}$ was found to be $g_c^{(np)}=1.008$ for both the classical and \textit{qasm} implementations. Similarly, the critical point for the case of $H_{sp}$ was found to be $g_c^{(sp)}=0.996$ for both the classical and \textit{qasm} implementations. Here we notice that although, the convergence for the data obtained from both the classical and $qasm$ implementations turns out to be the same for both $H_{np}$ and $H_{sp}$, such a perfect match appears to be somewhat coincidental. In Appendix~\ref{bsa}, we have explained the Bulirsch-Stoer algorithm which sets the criteria used to deduce these convergence results. The convergence plots have been added to the \textit{Supporting Information} section.

\section{Discussion and Outlook}
\label{discussion}

In this paper we have used the Finite-Size Scaling in Hilbert Space approach to calculate the critical point of the Quantum Rabi Model. We used the low-energy effective Hamiltonians for both the normal and superradiant phases respectively to show that the critical point is $g_c \approx 1$. The original FSS approach in which the truncation is done in the physical space has been widely used to calculate critical points and critical exponents since its inception. However that approach was not applicable to Quantum Phase Transitions which occur at a finite system size. With the rise in interest in QPTs occurring in these finite size systems, our approach provides a natural extension of the original FSS method to study such phase transitions. To our knowledge, this is the first time this approach has been used to study a QPT in a light-matter interaction system.

We have also provided a recipe for the implementation of this method on a universal quantum computer using the Quantum Restricted Boltzmann Machine algorithm. It was shown that results obtained from the classical gate simulation match those obtained from the IBM-Q's \textit{qasm} simulator. Such an implementation scales quadratically while the exact diagonalization scales cubically in the best case and exponentially in the worst case. Looking forward, we are interested in applying this approach to other QPTs such as the QPT in anisotropic QRM. We would also like to use our method to calculate the critical exponents in addition to the critical points in these phase transitions. It would also be interesting to see if this approach can be used to predict any new phase transition for some other non-integrable model.

Another very promising research direction is to implement the FSS method for phase transitions in classically intractable many-body models such as exotic electronic and magnetic systems. These include general quantum materials, for example where Coulomb potential leads to a gapped spectrum in energy, including in direct band-gap semiconductors in the thermodynamic limit. Conventionally speaking, it might be necessary to resort to the original finite-size scaling in the physical space approach for these systems since they exhibit criticality only in the limit $N \to \infty$. However, the ground state of an appropriately truncated Hamiltonian could be deduced using the QRBM algorithm as shown in the paper towards efficient implementation on a digital quantum simulator. A simile can also be drawn between a many-body bulk gap separating a continuum of excited states from the ground state manifold to the gapped Rabi model discussed in this paper. Such an approach can be useful in emergent topological systems, such as in Weyl semimetals, 1-D Kitaev spin chains, quantum spin liquids, and others, on which there is a tremendous explosion of interest \cite{Google, Kitaev, pollmann, Xiao, Wen, Zahid}. Topological phase transitions are devoid of any conventional order parameter and a quantum solution deriving from the approach outlines in this paper can help us bypass resource and scaling limitations of DMRG and exact diagonalization approaches to calculate the critical point and the critical exponents.

\begin{acknowledgement}

We acknowledge funding by the US Department of Energy, Office of Science, National Quantum Information Science Research Centers, Quantum Science Center.

\end{acknowledgement}

\begin{suppinfo}

Fig.~\ref{Errors} displays the enlarged error plots included in the insets of Fig.~\ref{QRBM_results}. Fig.~\ref{Convergence} shows the convergence plots for the data in Fig.~\ref{QRBM_results}.

\begin{figure}[H]
    \centering
    \includegraphics[width=1\textwidth]{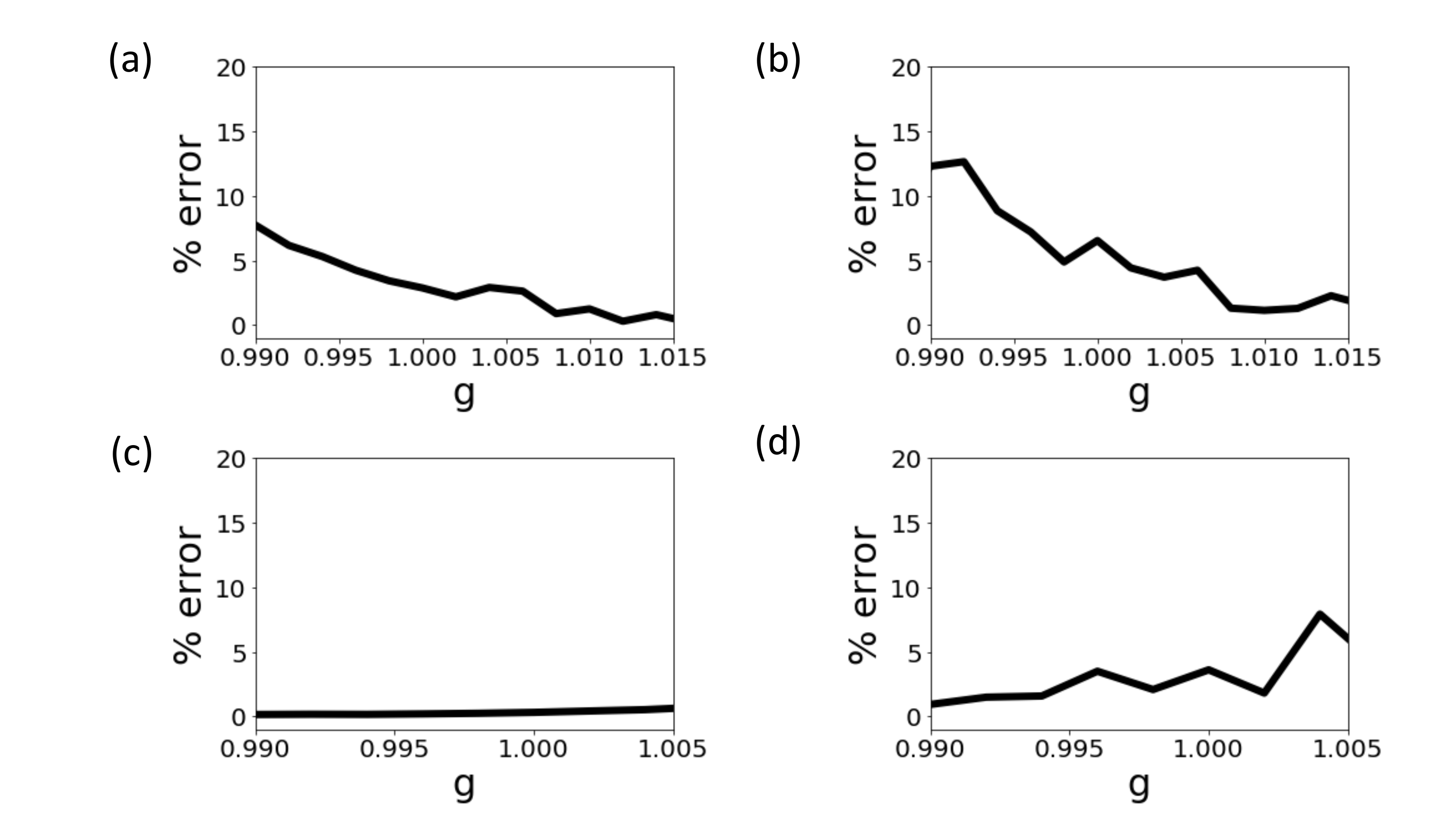}
    \caption{\textbf{Error plots from the insets of Fig~\ref{QRBM_results}.}} 
    \label{Errors}
\end{figure}

\begin{figure}[H]
    \centering
    \includegraphics[width=1\textwidth]{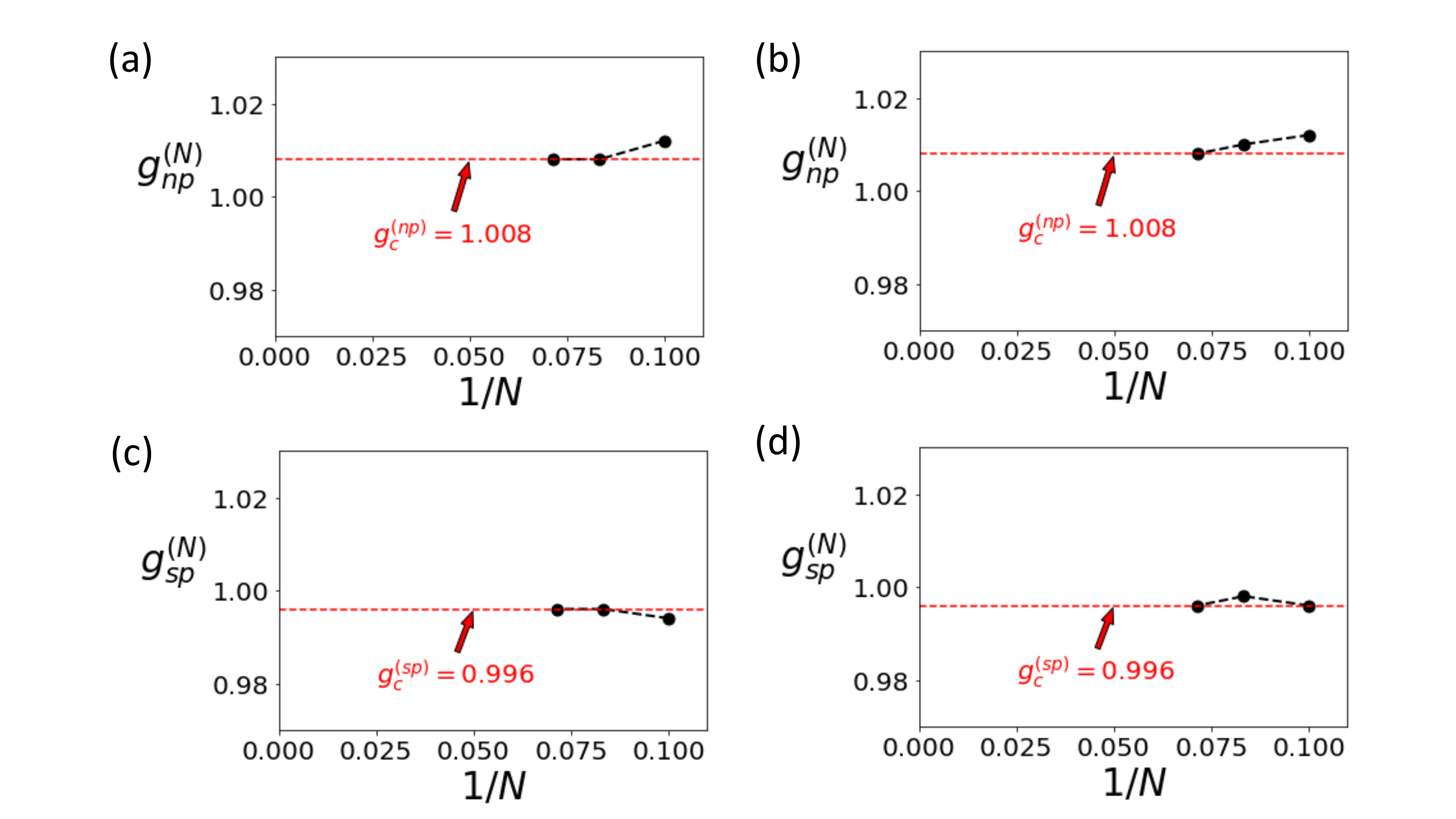}
    \caption{\textbf{Convergence diagrams for results in Fig~\ref{QRBM_results}.} (a), (b), (c), (d) correspond to convergence results for data in Fig.~\ref{QRBM_results}(a), (b), (c), (d) respectively. The same procedure was used as the one shown in Fig.~\ref{qrm_fss}(b) and (d).} 
    \label{Convergence}
\end{figure}

\end{suppinfo}

\begin{appendices}

\section{Bulirsch-Stoer Algorithm}
\label{bsa}

For $h_N=1/N$ where $N=0,1,2,\ldots$, the Bulirsch-Stoer algorithm can be used to find the limit of a function $T(h_N)$ as $N \to \infty$\cite{bsa1,bsa2}. For demonstration, consider that we only have $T(h_N)$ for $N=0,1,2,3$, then the following rows are computed successively, \\ \\
\begin{tabular}{>{$n=}l<{$\hspace{12pt}}*{13}{c}}
$0$    &&&&$T_0^{(0)}$&&$T_0^{(1)}$&&$T_0^{(2)}$&&$T_0^{(3)}$&&&\\
$1$    &&&&&$T_1^{(0)}$&&$T_1^{(1)}$&&$T_1^{(2)}$&&&&\\
$2$    &&&&&&$T_2^{(0)}$&&$T_2^{(1)}$&&&&&\\
$3$    &&&&&&&$T_3^{(0)}$&&&&&&
\end{tabular} \\ \\
using the following rules
\begin{align}
    T_{-1}^{(N)} &= 0 \\
    T_{0}^{(N)} &= T(h_N) \\
    T_{m \geq 1}^{(N)} &= T_{m-1}^{(N+1)} + (T_{m-1}^{(N+1)} - T_{m-1}^{(N)}) \Bigg[\Bigg(\frac{h_N}{h_{N+m}}\Bigg)^\omega \Bigg(1 - \frac{T_{m-1}^{(N+1)} - T_{m-1}^{(N)}}{T_{m-1}^{(N+1)} - T_{m-2}^{(N+1)}}\Bigg) - 1 \Bigg]^{-1}
\end{align}
where $\omega$ is a free parameter determined by minimizing $\varepsilon_m^{(i)}=\abs{T_{m}^{(i+1)} - T_{m}^{(i)}}$. The final answer is $T_3^{(0)}.$

\end{appendices}

\bibliography{references}

\end{document}